\documentclass[12pt,amsfonts]{iopart}

\usepackage{epsfig,psfrag}
\usepackage{latexsym}
\usepackage{graphicx}
\usepackage{axodraw}

\def\beq{\begin{eqnarray}}
\def\eeq{\end{eqnarray}}
\def\bea{\begin{eqnarray}}
\def\eea{\end{eqnarray}}
\def\det{{\rm det}}
\newcommand\eqn[1]{(\ref{#1})}      % parentheses around the LaTex "ref" macro
\newcommand{\nn}{\nonumber}
\newcommand{\reals}{\mbox{${\rm I\!R }$}}

\def\phic{\Phi_{\rm cl}}

\usepackage{color}

\def\firstorder{\begin{picture}(10,10)(10,10)
\CArc(15,15)(10,0,180)
\CArc(15,15)(10,180,360)
\Vertex(5,15)2
\end{picture}}

\def\secondorder{\begin{picture}(10,10)(10,10)
\CArc(15,15)(10,0,180)
\CArc(15,15)(10,180,360)
\Vertex(5,15)2
\Vertex(25,15)2
\end{picture}}

\def\thirdorder{\begin{picture}(10,10)(10,10)
\CArc(15,15)(10,0,180)
\CArc(15,15)(10,180,360)
\Vertex(15,25)2
\Vertex(24,12)2
\Vertex(6,12)2
\end{picture}}

\begin{document}

\title[Functional Determinants in Quantum Field Theory]{Functional Determinants in Quantum Field Theory\footnote{Plenary talk at QTS5, Quantum Theories and Symmetries, Valladolid, August 2007.}}
\author{Gerald V. Dunne}
%\email{dunne@phys.uconn.edu}
\address{Institut f\"ur Theoretische Physik,  Universit\"at Heidelberg, 
69120 Heidelberg, Germany\\
Department of Physics, University of Connecticut, Storrs, CT 06269, USA}

\begin{abstract}
Functional determinants of differential operators play a prominent role in theoretical and mathematical physics, and in particular in quantum field theory.  They are, however, difficult to compute  in non-trivial cases.  For one dimensional problems, a classical result of Gel'fand and Yaglom dramatically simplifies the problem so that the functional determinant can be computed without computing the spectrum of eigenvalues. Here I report  recent progress in extending this approach to higher dimensions (i.e., functional determinants of {\it partial} differential operators), with applications in quantum field theory.
\end{abstract}

%\maketitle

\section{Introduction : why are determinants important in QFT?}
\label{sec-intro}

This talk considers the question: {\it what is the determinant of a partial differential operator, and how might one compute it?} 

The physical motivation for this question comes from several areas of physics:
\begin{itemize}
\item
Effective actions, grand canonical potentials: These appear in relativistic and non-relativistic many-body theory and quantum field theory (QFT), and are generically given by an expression of the form
$\Omega =\tr\,\ln\, G=\ln\, \det\, G$, where $G$ is a Green's function, and the trace is over an infinite set of quantum states.

\item
Tunneling and semiclassical physics: tunneling and nucleation processes can be analyzed in a semiclassical approach, wherein the leading exponential contribution involves the  action evaluated on a classical solution, while the next-to-leading prefactor is the determinant of the fluctuation operator describing quantum fluctuations about the classical solution:
\bea
\int {\mathcal D}\phi\, e^{-S[\phi]}\approx \frac{e^{-S[\phi_{cl}]}}{\sqrt{{\rm det(fluctuation\,\,operator)}}}
\eea

\item
Gap equations: in many-body theory and relativistic QFT, one can find ground states by solving self-consistent Hartree-Fock or Schwinger-Dyson equations, varying the grand canonical potential or effective action with respect to a condensate field. Symbolically, the problem is of the form
\bea
\sigma(x)=\frac{\delta}{\delta \sigma(x)}\ln\,\det\left[{\mathcal D} +\sigma(x)\right]
\eea
where ${\mathcal D}$ is some differential operator. Clearly, to be able to vary the determinant efficiently, one needs a clever way to evaluate the determinant.

\item
Lattice gauge theory: integrating out quarks leads to fermion determinant factors
\bea
\int {\mathcal D}\psi \, {\mathcal D}\bar{\psi} \, e^{-\int d^4x\,\bar{\psi}\left[ D\hskip -5pt /  -m\right] \psi}
=\det\left(D\hskip -8pt /  -m\right)
\eea
which are needed for dynamical fermion computations in QCD. There are still fundamental challenges in evaluating such determinants within the Monte Carlo approach to lattice gauge theory, especially at finite density and temperature.

\item
Faddeev-Popov determinants: the determinant of the Faddeev-Popov operator plays an important role in  gauge fixing and confinement.

\item
Mathematical physics: 
the determinant of a partial differential operator, such as a Dirac or Klein-Gordon operator, encodes interesting spectral information.

\end{itemize}

\section{What we know}

Exact results for functional determinants exist only for certain special cases. For example, the gauge theory effective action [the logarithm of the determinant of a Dirac or Klein-Gordon operator] for background fields with (covariantly) constant field strength can be expressed as simple proper-time integrals, building on the seminal work of Heisenberg and Euler \cite{he,schwinger,gvd}. Mathematically, these expressions are closely related to the multiple zeta and gamma functions \cite{ruijsenaars}. This constant curvature case is soluble because we can choose a Fock-Schwinger gauge with $A_\mu=-\frac{1}{2}F_{\mu\nu}\,x_\nu$, in which case the spectral problem decomposes into a set of harmonic oscillator problems, for which the spectral information is very simple. Similarly, in gravitational problems, for special spaces such as harmonic spaces or symmetric spaces, the spectral problem is sufficiently separable and simple that the determinant can be evaluated \cite{camporesi}. For example, the determinant of Dirac or Klein-Gordon operators on spheres, cones and tori are known in quite explicit form \cite{camporesi,dowker}. Remarkably, this type of analysis can also be extended to general Riemann surfaces \cite{dhoker, sarnak}, with important applications to worldsheet problems in string theory and with beautiful applications in number theory and differential geometry.

Things become more interesting when including general non-constant curvatures [either gauge or gravitational]. Then there are a few especially symmetric cases that are still soluble, but they are all one-dimensional. For example, in gauge theory, if the field strength points in a particular fixed direction, and has a magnitude that is a function of only one coordinate [say $x_\mu$], and is of the form $F\,{\rm sech}^2\left(k \, x_\mu\right)$, then the corresponding [Dirac or Klein-Gordon] spectral problem is hypergeometric, and the determinant can be computed explicitly. Also, for massless quarks in an instanton background, the fermion determinant is known.  Such a computation was pioneered by 't Hooft for the single $su(2)$ instanton \cite{thooft}, but has since been generalized to instantons of full ADHM form \cite{osborn}, to finite temperature \cite{gross}, and to calorons with nontrivial holonomy \cite{diakonov}. These computations all rely on the fact that the Green's functions for massless particles in an instanton background are known explicitly. The generalization to nonzero masses is nontrivial, and is discussed below in Section \ref{idet-sec}.

Some general bounds are known, based on lattice regularizations and generalizations of Kato's inequality for Dirac and Klein-Gordon operators. For example \cite{seiler,vafawitten}, 
\bea
\left| \frac{
{\rm det}\left[m^2-D\hskip -7pt /\,^2\right]}
{{\rm det}\left[m^2-\partial\hskip -6pt /\,^2\right]}\right|
\leq 1 \qquad ; \qquad 
\left| \frac{
{\rm det}\left[m^2-D_\mu^2\right]}
{{\rm det}\left[m^2-\partial_\mu^2\right]}\right|
\geq 1 \quad .
\eea 
While very general, these bounds are still quite weak, and do not contain information about the phase of the determinant, which can be physically significant. Tighter bounds have been found recently in QED \cite{fry}, but general results are still relatively limited.

For Schr\"odinger operators with general potentials, or Dirac/Klein-Gordon operators with arbitrary gauge and/or gravitational backgrounds, we rely heavily on approximate methods. One such approximation is the heavy mass expansion. Define
\bea
\ln\,\det \left[m^2+{\mathcal D}\right]=\tr\,\ln \left[m^2+{\mathcal D}\right] =-\int_0^\infty \frac{ds}{s}\, e^{-m^2 s}\,
\tr\left\{e^{-s\, {\mathcal D}}\right\} 
\eea
Then an inverse mass expansion follows (after renormalization) from the small $s$ asymptotic expansion of the heat kernel operator \cite{vassilevich}
\bea
\tr\left\{e^{-s\, {\mathcal D}}\right\} \sim \frac{1}{(4\pi\, s)^{d/2}}
\sum_{k=0}^\infty s^k\, a_k\left[ {\mathcal D}\right] \qquad .
\eea
Here the $a_k\left[ {\mathcal D}\right]$ are known functionals of the potentials appearing in ${\mathcal D}$. Similarly, a derivative expansion can be derived by expanding the heat kernel trace $\tr\left\{e^{-s\, {\mathcal D}}\right\} $ about the soluble constant background case, in powers of derivatives of the background. This corresponds to resumming all non-derivative terms in the inverse mass expansion. While very general, these expansions are asymptotic, with higher terms becoming rapidly unwieldy, and so have a somewhat limited range of application.

Another  approach is to define the  functional determinant via a zeta function 
\bea
\zeta(s)\equiv \sum_\lambda \frac{1}{\lambda^s} \qquad ,
\eea
where the sum is over the spectrum of the relevant differential operator. Then, by the formal manipulations $\zeta^\prime(0)=-\sum_\lambda \ln\,\lambda=-\ln\left(\prod_{\lambda}\lambda\right)$,  one {\it defines} the determinant as
\bea
{\rm det}=e^{-\zeta^\prime(0)} \qquad .
\eea
The problem becomes one of analytically continuing $\zeta(s)$ from the region in which it converges [typically  $\Re (s) > d/2$] to the neighbourhood of $s=0$. For example, in a radially separable problem, one might estimate the spectrum using WKB phase shifts [and the corresponding spectral function $\rho(k)=\frac{1}{\pi}\frac{d\delta}{d k}$], thereby obtaining information about the zeta function.
The zeta function approach is complementary to the so-called {\it replica method}, where one defines the logarithm of an operator by considering {\it positive} integer powers $n$ of the operator, and then analytically continues to $n=0$. There are also non-trivial interesting one dimensional quantum mechanical problems for which the determinant can be computed from the zeta function \cite{voros}.

There are several special features of one-dimensional problems that deserve mention. First, the issue of renormalization does not really enter in 1d. Second, the theory of inverse scattering allows us to characterize a potential in terms of the associated scattering data. This is especially useful for solving variational and gap equations, as the variational problem can then be reformulated in terms of the scattering data, which is more direct. Beautiful applications of this idea occur in interacting 1+1 dimensional QFT's such as the Gross-Neveu model, Sine-Gordon model and their generalizations \cite{dhn,thies}. [Inverse scattering has been generalized in higher dimensions for radial problems \cite{avan}]. A third special feature of 1d problems is the Gel'fand Yaglom theorem \cite{gy,levit}, which provides a way to compute the determinant of an ordinary differential operator without computing its eigenvalues. The basic idea is that in 1d one can define the Green's function without using an eigenfunction expansion, simply from the product of two independent solutions.

\section{Gel'fand-Yaglom theorem: ordinary differential operators}

The simplest form of this result can be stated as follows \cite{gy,levit,coleman,forman,kappeler,kleinert}. Consider a Schr\"odinger operator on the interval $x\in [0, L]$, with Dirichlet boundary conditions:
\bea
\left[-\frac{d^2}{dx^2}+V(x)\right]\psi(x)=\lambda \, \psi(x)\qquad ; \qquad \psi(0)=\psi(L)=0
\eea
Then to compute the determinant it is in fact not necessary to compute the infinite discrete set of eigenvalues $\left\{\lambda_1, \lambda_2, \dots\right\}$. Instead, solve the related {\it initial value problem}
\bea
\left[-\frac{d^2}{dx^2}+V(x)\right]\phi(x)=0 \qquad ; \qquad \phi(0)=0\qquad ; \qquad \phi^\prime(0)=1
\label{ivp}
\eea
Then
\bea
\det  \left[-\frac{d^2}{dx^2}+V(x)\right]=\phi(L)\qquad .
\label{gy1}
\eea
It is worth pausing to appreciate the simplicity of this result, as well as its practical utility. It is straightforward to implement numerically. The determinant follows, without having to compute any eigenvalues, let alone multiply them all together!

Actually, \eqn{gy1} is not quite right, as we can only define the {\it ratio} of two determinants \cite{simon}, and \eqn{gy1} should be understood in this sense. Often in physical applications one is computing the determinant {\it relative to} the corresponding determinant for the free operator, for which the initial value problem is usually known in closed form [in this case it is numerically better to solve directly the initial value problem for the {\it ratio}]. As an illustration, consider the following simple example of the massive Helmholtz operator $\left[-\frac{d^2}{dx^2}+m^2\right]$. The Dirichlet  spectrum is  $\lambda_n=m^2+\left(\frac{n\,\pi}{L}\right)^2$, so from the eigenvalues:
\bea
\hskip -1cm \frac{\det  \left[-\frac{d^2}{dx^2}+m^2\right]}{\det  \left[-\frac{d^2}{dx^2}\right]} =
\prod_{n=1}^\infty\frac{\left[m^2+\left(\frac{n\,\pi}{L}\right)^2\right]}{\left[\left(\frac{n\,\pi}{L}\right)^2\right]}=
\prod_{n=1}^\infty\left[1+\left(\frac{mL}{n\,\pi}\right)^2\right]=\frac{\sinh(m\, L)}{m\, L}
\label{1dh-1}
\eea
On the other hand, if we use the Gel'fand-Yaglom result, we solve the initial value problems: $\left[-\frac{d^2}{dx^2}+m^2\right]\phi=0$, to find $\phi(x)=\frac{\sinh(m\, x)}{m}$; and $-\frac{d^2}{dx^2} \phi_0=0$, to find $\phi_0(x)=x$ for the massless operator. Then \eqn{gy1} implies 
\bea
\hskip -1cm \frac{\det  \left[-\frac{d^2}{dx^2}+m^2\right]}{\det  \left[-\frac{d^2}{dx^2}\right]} =\frac{\phi(L)}{\phi_0(L)}=
\frac{\sinh(m\, L)}{m\, L}
\label{1dh-2}
\eea
These clearly agree, but the point is that if $m^2$ were replaced by a nontrivial potential $V(x)$, the first approach, from the eigenvalues, would be extremely difficult, while the Gel'fand-Yaglom approach is still easy.
As another example, consider the P\"oschl-Teller system with $H=[-\frac{d^2}{dx^2}+m^2-j(j+1)\,{\rm sech}^2\,x]$, with integer $j$. This is a reflectionless potential with $j$ bound states at $E=m^2-l^2$, $(l=1,\dots j$), and phase shift $\delta(k)=2\sum_{l=1}^j {\rm arctan}(l/k)$. The determinant follows directly from the spectrum as
\bea
\hskip -2cm \ln\left(\frac{\det [-\frac{d^2}{dx^2}+m^2-j(j+1){\rm sech}^2\,x]}{\det[-\frac{d^2}{dx^2}+m^2]} \right)&=&\sum_{l=1}^j\ln(m^2-l^2)+\int_0^\infty \frac{dk}{\pi}\frac{d\delta}{dk}\ln(m^2+k^2)\nonumber\\
&=&\ln\frac{\Gamma(m)\Gamma(m+1)}{\Gamma(m-j)\Gamma(m+j+1)}
\eea
On the other hand, the solutions to the Gel'fand-Yaglom initial value problems \eqn{ivp} are
\bea
\hskip -2cm \phi(x)&=&\frac{1}{4^m}
\frac{\Gamma\left(\frac{j-m+2}{2}\right)\Gamma\left(\frac{j-m+1}{2}\right)}{\Gamma\left(\frac{j+m+2}{2}\right)\Gamma\left(\frac{j+m+1}{2}\right)}\left(P_j^m(-{\rm th}\, L)\, Q_j^m({\rm th} \, x)- Q_j^m(-{\rm th} \, L) P_j^m({\rm th}\, x)\right)\nonumber\\
\hskip -2cm \phi_{(j=0)}(x)&=&\frac{1}{m}\left(\cosh(m\, L)\, \sinh(m\, x)+ \sinh(m\, L)\, \cosh(m\, x)\right)
\eea
on the interval $x\in [-L, L]$. The $P_j^m$ and $Q_j^m$ are Legendre functions. Then 
\bea
\hskip -2cm \frac{\det [-\frac{d^2}{dx^2}+m^2-j(j+1){\rm sech}^2\,x]}{\det[-\frac{d^2}{dx^2}+m^2]} =\lim_{L\to\infty} \frac{\phi(L)}{\phi_{(j=0)}(L)}=\frac{\Gamma(m)\Gamma(m+1)}{\Gamma(m-j)\Gamma(m+j+1)}
\eea
Similarly, one can define a general reflectionless potential with $N$ bound states at $E=m^2-\kappa_i^2$, $(i=1,\dots N$), by $V(x)=m^2-2\frac{d^2}{dx^2}\ln\,\det \, A(x)$ where $A(x)$ is the $N\times N$ matrix $A_{ij}=\delta_{ij}+\frac{c_i\, c_j}{\kappa_i+\kappa_j}e^{-(\kappa_i+\kappa_j)x}$. The $c_i$ are (constant) moduli which affect the shape of the potential, but not the spectrum. From the phase shift $\delta(k)=2\sum_{i=1}^N {\rm arctan}(\kappa_i/k)$, one finds the determinant
\bea
 \frac{\det [-\frac{d^2}{dx^2}+m^2+V(x)]}{\det[-\frac{d^2}{dx^2}+m^2]} =\exp\left[-2\sum_{i=1}^N {\rm arctanh}\left(\frac{m}{\kappa_i}\right)\right]
\eea
It is instructive to verify this numerically for various potentials using \eqn{ivp} and \eqn{gy1}.

The basic result \eqn{gy1} generalizes in several ways. First, it generalizes to other boundary conditions \cite{forman,kleinert,kirsten}, characterized by two $2\times 2$ matrices $M$ and $N$ as
\bea
M\left(\matrix{\psi(0)\cr\psi^\prime(0)}\right)+N \left(\matrix{\psi(L)\cr\psi^\prime(L)}\right)=0
\eea
Construct two independent solutions $u_{(1),(2)}$  of $\left[-\frac{d^2}{dx^2}+V(x)\right]\phi(x)=0$ such that
\bea
u_{(1)}(0)=1 \qquad &;& \qquad u^\prime_{(1)}(0)=0\nonumber\\
u_{(2)}(0)=0 \qquad &;& \qquad u^\prime_{(2)}(0)=1
\eea
Then the infinite dimensional determinant reduces to a simple $2\times 2$ determinant
\bea
\det  \left[-\frac{d^2}{dx^2}+V(x)\right]= \det_{2\times 2}\left[M+N \left(\matrix{u_{(1)}(L)&u_{(2)}(L)\cr
u^\prime_{(1)}(L)&u^\prime_{(2)}(L)}\right)\right]
\eea
For example, 
\bea
\hskip -1cm {\rm Dirichlet}: M=\left(\matrix{1&0\cr 0&0}\right), N=\left(\matrix{0&0\cr 1&0}\right) &\Rightarrow& \det  =u_{(2)}(L) \nonumber\\
\hskip -1cm {\rm Neumann}: M=\left(\matrix{0&0\cr 0&1}\right), N=\left(\matrix{0&1\cr 0&0}\right) &\Rightarrow& \det  =u_{(1)}^\prime(L) \nonumber\\
\hskip -1cm{\rm Periodic\,\,(P)}: M={\bf 1}, N=-{\bf 1} &\Rightarrow& \det  =2-\left(u_{(1)}(L)+u^\prime_{(2)}(L)\right)\nonumber\\
\hskip -1cm{\rm Antiperiodic\,\, (AP)}: M={\bf 1}, N={\bf 1}&\Rightarrow& \det  =2+\left(u_{(1)}(L)+u^\prime_{(2)}(L)\right)\nonumber
\eea
In the P/AP cases we recognize $(u_{(1)}(L)+u^\prime_{(2)}(L))$ as the Floquet-Bloch  {\it discriminant}, which takes values $\pm 2$ for P/AP eigenfunctions.
Second, these results generalize straightforwardly to coupled systems of ODE's \cite{forman,kirsten}, and also to linear ODE's of any order \cite{dym}. Third, they generalize to Sturm-Liouville problems \cite{kirsten}. For example, consider the radial operator in $d$ dim., with Dirichlet b.c.'s
\bea {\mathcal M}_{(l)}\equiv
-\frac{d^2}{dr^2}+\frac{\left(l+\frac{d-3}{2}\right)\left(l+\frac{d-1}{2}\right)}{r^2}+V(r)\quad
. \label{radial-op} 
\eea 
Define the function $\phi_{(l)}(r)$ as the solution to the radial initial value problem 
\bea
\left[{\mathcal M}_{(l)}+m^2\right] \phi_{(l)} (r) =0\qquad ; \quad
\phi_{(l)}(r)\sim r^{l+(d-1)/2} \quad,\quad  {\rm as}\quad r\to
0 \quad .
\label{phi-l} 
\eea 
Then (this should also be understood as applying to {\it ratios} of determinants)
\bea
\det  \left[{\mathcal M}_{(l)}+V(r)\right]=\phi_{(l)}(R)\qquad .
\label{gyr}
\eea
As an explicit example, consider the 2d radial Helmholtz problem, with eigenvalues in terms of the zeros $j_{(l),n}$ of the $J_l$ Bessel  function: $\lambda_{(l), n}=m^2+\left(\frac{j_{(l),n}}{R}\right)^2$. Then
\bea
\hskip -1cm \frac{\det  \left[{\mathcal M}_{(l)}+m^2\right]}{\det  \left[{\mathcal M}_{(l)} \right]} =
\prod_{n=1}^\infty\frac{\left[m^2+\left(\frac{j_{(l),n}}{R}\right)^2 \right]}{\left[\left(\frac{j_{(l),n}}{R}\right)^2\right]} =
\prod_{n=1}^\infty\left[1+\left(\frac{m\, R}{j_{(l),n}}\right)^2\right]
\label{radial-eig}
\eea
On the other hand, if we use the Gel'fand-Yaglom result \eqn{gyr}, we readily solve the relevant initial value problems to find $\phi_{(l)}$ in terms of the modified Bessel function $I_l$:
\bea
\frac{\phi_{(l)}(R)}{\phi^{\rm free}_{(l)}(R)}=\frac{l!\, I_l(m\, R)}{\left(\frac{m\, R}{2}\right)^l} \qquad .
\label{radial-bessel}
\eea
The agreement between \eqn{radial-eig} and \eqn{radial-bessel}  expresses the product formula for $I_l(x)$.
%$I_l(x)=\frac{x^l}{2^l\, l!} \prod_{n=1}^\infty \left[1+(\frac{x}{j_{(l),n}})^2\right]$.

\section{Gel'fand-Yaglom theorem: partial differential operators?}

In generalizing from {\it ordinary} to {\it partial} differential operators, it is natural to consider first the case of {\it separable} operators. But the na\"ive extension fails. Formally, we simply take a product over the radial determinants [each of which is evaluated as above] for all possible angular momenta, with the appropriate degeneracies. However, this product diverges \cite{forman}. For example, for the 2d radial Helmholtz system \eqn{radial-bessel}
\bea
\ln\left( \frac{\det  \left[{\mathcal M}_{(l)}+m^2\right]}{\det  \left[{\mathcal M}_{(l)} \right]}\right) =\ln\left( \frac{l!\,2^l\, \, I_l(m\, R)}{(m\, R)^l} \right) \sim O\left(\frac{1}{l}\right)\quad, \quad l\to\infty
\eea
The degeneracy factor is $1$ for $l=0$, and $2$ for all $l\geq 1$, so the angular momentum sum over $l$ diverges. This should not be so surprising from a physical point of view, as we expect to require renormalization in dimension higher than 1.
To formulate the problem more precisely, consider the radially separable 
operators
$
{\mathcal M}=-\Delta+V(r)$, and ${\mathcal M}^{\rm free}=-\Delta
$,
where $\Delta$ is the Laplace operator in $\reals^d$, and $V(r)$
is a radial potential vanishing at infinity as $r^{-2-\epsilon}$
for $d=2$ and $d=3$, and as $r^{-4-\epsilon}$ for $d=4$. 
Since $V=V(r)$,
we can separate variables and consider the 
Schr\"odinger-like radial operator ${\mathcal M}_{(l)}$ in \eqn{radial-op}. For dimension
$d\geq 2$, the radial eigenfunctions $\psi_{(l)}$ have degeneracy 
\bea {\rm deg}(l ; d)\equiv
\frac{(2l+d-2)(l+d-3)!}{l!(d-2)!} \quad . 
\label{deg} 
\eea
Formally, we write
\bea \ln\left(\frac{\det
\left[{\mathcal M}+m^2\right]}{\det \left[{\mathcal M}^{\rm
free}+m^2\right]}\right)=\sum_{l=0}^\infty {\rm deg}(l; d)
\ln\left(\frac{\det \left[{\mathcal M}_{(l)}+m^2\right]}{\det
\left[{\mathcal M}^{\rm free}_{(l)}+m^2\right]}\right) \quad .
\label{formal-sum} 
\eea 
Each term in the sum is computed straightforwardly using \eqn{gyr}, but the  $l$ sum is divergent.
However, this divergence can be understood physically, leading to  a finite and renormalized determinant
ratio \cite{dk}:
\bea
 \hskip -2cm \ln\left(\frac{\det [{\mathcal M}+m^2]}{\det [{\mathcal M}^{\rm free}+m^2]}\right)\Bigg |_{d=2}
& =& \ln \left(\frac{\phi_{(0)}(\infty)}{\phi^{\rm free}_{(0)}(\infty)}\right)+
\sum_{l=1}^\infty 2 \left\{\ln \left(\frac{\phi_{(l)}(\infty)}{\phi^{\rm free}_{(l)}(\infty)}\right)-\frac{\int_0^\infty dr\, r V(r)}{2l}\right\} \nn\\
&&  +\int_0^\infty dr\, r\, V \left[\ln\left(\frac{\mu r}{2}\right)+\gamma \right] 
\label{2d-result}  \nn\\
\hskip -2cm  \ln\left(\frac{\det [{\mathcal M}+m^2]}{\det [{\mathcal M}^{\rm free}+m^2]}\right)\Bigg |_{d=3}
& =& \sum_{l=0}^\infty \left(2 l+ 1 \right) \left\{
\ln \left(\frac{\phi_{(l)}(\infty)}{\phi^{\rm free}_{(l)}(\infty)}\right)-\frac{\int_0^\infty dr\, r V(r)}{2(l+\frac{1}{2})}\right\}
\label{3d-result} \nn\\
\hskip -2cm  \ln\left(\frac{\det [{\mathcal M}+m^2]}{\det [{\mathcal M}^{\rm free}+m^2]}\right)\Bigg |_{d=4}&  =& \hskip -2cm \nn\\
& &\hskip-3cm \sum_{l=0}^\infty
\left(l+1\right)^2 \left\{
\ln \left(\frac{\phi_{(l)}(\infty)}{\phi^{\rm free}_{(l)}(\infty)}\right)-\frac{\int_0^\infty dr\, r V(r)}{2(l+1)}+\frac{\int_0^\infty dr\, r^3 V(V+2m^2)}{8(l+1)^3}\right\} \nn\\
&\hskip -2cm &\hskip -2cm  -\frac{1}{8}\int_0^\infty dr\, r^3\, V(V+2m^2) \left[\ln\left(\frac{\mu r}{2}\right)+\gamma+1\right] \quad .
\label{4d-result}
\eea
Here $\gamma$ is Euler's constant, and $\mu$ is a renormalization scale, which is essential for physical applications. A conventional renormalization choice is to take $\mu=m$ in \eqn{4d-result}. In each of \eqn{4d-result}, the sum over $l$ is convergent once
the indicated subtractions are made. The function $\phi_{(l)}(r)$ is defined in \eqn{phi-l}.

Notice that the results \eqn{4d-result} state
once again that the determinant is determined by the boundary
values of solutions of $\left[{\mathcal M}+m^2\right]\phi=0$, with
the only additional information being a finite number of integrals
involving the potential $V(r)$. We also stress the computational
simplicity of \eqn{4d-result}; no phase shifts or eigenvalues need be computed to evaluate the renormalized determinant. 

These results have been derived \cite{dk} using the zeta function formalism.
Scattering theory \cite{taylor} and standard contour manipulations \cite{kirsten,kirstenbook} yield an integral representation [valid for $\Re (s) > d/2$] of the zeta function in terms of the Jost function $f_l(ik)$:
\begin{eqnarray}
\zeta(s)  ={\sin \pi s\over \pi} \sum_{l=0}^{\infty}{\rm deg}(l; d)
\, \int\limits_{m}^{\infty}dk\,\,
[k^2-m^2]^{-s}~\frac{\partial}{\partial k}\ln  f_l (ik)\quad .
\label{zeta-jost}
\end{eqnarray}
The technical problem is the analytic continuation of \eqn{zeta-jost} to a neighborhood about $s=0$. 
This analytic
continuation relies on the uniform asymptotic behavior of the Jost
function $f_l (i k)$, which follows from standard results in scattering theory \cite{taylor}:
\begin{eqnarray}
\hskip -2.5 cm \ln f_l (ik) = \int\limits_0^{\infty}dr\, r\, V(r)
K_{\nu} (kr) I_{\nu} (kr)-\int\limits_0^{\infty}dr\,r\, V(r)
K_{\nu}^2 (kr) \int\limits_0^rdr'\, r' \,
 V(r') I_{\nu}^2 (kr')+ {\cal O} (V^3) \nn
\label{jost-expansion}
\end{eqnarray}
where $\nu\equiv l+\frac{d}{2}-1$. Then the calculation of the
asymptotics of the Jost function reduces to the known uniform asymptotics
of the modified Bessel functions $K_\nu$ and $I_\nu$. 
Using these asymptotics, we can {\it define} $\ln f_l^{asym}(ik)$ as the $O(V)$ and  $O(V^2)$ parts of this uniform asymptotic expansion, and then
by construction, 
\begin{eqnarray}
 \zeta_f' (0) &\equiv & -\sum_{l=0}^{\infty}{\rm deg}(l; d)
       \left[\ln f_l (i m) -\ln f_l^{asym} (i m)
\right] \quad .
\label {zetaprime-f}
\end{eqnarray}
is now convergent. Adding back the subtracted terms in a dimensional regularization scheme, leads \cite{dk} to the results in \eqn{4d-result}.

\section{Comparison With Feynman Diagram Approach}
\label{sec-feynman}

These determinants can also be derived in a Feynman diagrammatic approach, where
the perturbative expansion in powers of the potential $V$ is 
\bea
\ln\left(\frac{\det \left[{\mathcal M}+m^2\right]}{\det \left[{\mathcal M}^{\rm free}+m^2\right]}\right) & \equiv& \sum_{k=1}^\infty \frac{(-1)^{k+1}}{k}\, A^{(k)}\nn\\
&=&\quad
\firstorder\quad -\frac{1}{2} \quad
\secondorder \quad +\frac{1}{3} \quad
\thirdorder\quad +\dots\qquad ,
\label{feynman}
\eea
Here the solid dots denote insertions of the potential $V$.
If the potential is radial, then with dimensional regularization one can express the renormalized 4d determinant as \cite{baacke}
\bea
\hskip -2.5cm \left[\ln\left(\frac{\det \left[{\mathcal M}+m^2\right]}{\det \left[{\mathcal M}^{\rm free}+m^2\right]}\right)\right]_{d=4} &=& \sum_{l=0}^\infty  (l+1)^2\left\{ \ln f_l(i m) - \int_0^\infty dr\,r \, V(r) K_{l+1}(m r)\, I_{l+1}(m r)\right.\nonumber\\
&&\left. \hskip -3cm +\int_0^\infty dr\, r\, V(r) K_{l+1}^2(m r) \int_0^r dr^\prime r^\prime V(r^\prime) I_{l+1}^2(m r^\prime)\right\}+A^{(1)}_{\rm fin}-\frac{1}{2} A^{(2)}_{\rm fin} 
\label{baacke-split}
\eea
where the finite contributions from the first and second order Feynman diagrams in the $\overline{MS}$ scheme are \cite{baacke} (note the small typo in equation (4.32) of \cite{baacke}):
\bea
\hskip -1cm A_{\rm fin}^{(1)}&=&-\frac{m^2}{8}\int_0^\infty dr\, r^3\, V(r)\nn\\
\hskip -1cm  A_{\rm fin}^{(2)}&=&\frac{1}{128\pi^4}\int_0^\infty dq\, q^3\, \left| \tilde{V}(q)\right|^2\left[2-\frac{\sqrt{4m^2+q^2}}{q}\, \ln\left(\frac{\sqrt{4m^2+q^2}+q}{\sqrt{4m^2+q^2}-q}\right)\right] 
\label{a-finite}
\eea
Here $\tilde{V}(q)$ is the 4d Fourier transform of the radial potential $V(r)$. These expressions \eqn{baacke-split} and \eqn{a-finite} look quite different from \eqn{4d-result}, but in fact they can be shown to be completely equivalent, using some Bessel identities \cite{dk}.
However, it is clear that \eqn{4d-result} is much easier to evaluate, especially if (as often happens) $V(r)$ is known only numerically.

\section{Removing Zero Modes}
\label{zero-modes}

In certain quantum field theory applications the determinant may have zero modes, and correspondingly one is actually interested in computing the determinant with these zero modes removed, together with a collective coordinate factor \cite{coleman}. For example, consider the case of a self-interacting scalar field theory in $d$ dim Euclidean space with Euclidean action
$
S[\Phi]=\int d^d x\left[ \partial_\mu \Phi \partial_\mu \Phi +U(\Phi)\right]
$,
where there is a [radially symmetric] classical solution $\Phi_{cl}(r)$ solving $
\Delta \Phi=\frac{d U}{d\Phi}$, with $
\Phi^\prime (0)=0$, and 
$\Phi \to 0$  as $r\to \infty$.
The fluctuation operator about this classical solution is a radial operator, with
$V(r)=\left(\left[\frac{d^2 U}{d \Phi^2}\right]_{\Phi=\Phi_{cl}(r)} -m^2\right)$.
In the $l=1$ sector the corresponding radial operator ${\mathcal M}_{(l=1)}$ has a $d$-fold degenerate zero mode associated with translational invariance:
\bea
\psi_{\rm zero}(r)&=&\frac{\Phi_{cl}^\prime (r)}{\Phi^{\prime\prime}(0)} \quad.
\label{zero-mode}
\eea
Including the collective coordinate contribution \cite{coleman} due to translational invariance, the required determinant factor is [the prime denotes  exclusion of zero mode(s)]
\bea
\left(\frac{S[\Phi_{cl}]}{2\pi}\right)^{d/2} \left(\frac{\det^\prime  \left[{\mathcal M}_{(l=1)}+m^2\right]}{\det \left[{\mathcal M}^{\rm free}_{(l=1)}+m^2\right]} \right)^{-1/2} \qquad .
\label{coll}
\eea
Since the determinant is expressed in terms of a solution to an ODE, we can use simple ODE theory to show that (generalizing the $d=1$ \cite{mckane} and $d=4$  \cite{dunnemin} analyses)
\bea
\frac{\det^\prime  \left[{\mathcal M}_{(l=1)}+m^2\right]}{\det \left[{\mathcal M}^{\rm free}_{(l=1)}+m^2\right]}=\left[\frac{\int_0^\infty dr\, r^{d-1}\, \psi^2_{\rm zero}(r)}{\lim_{R\to \infty}\left(-2 R^{d-1} \psi_{\rm zero}^\prime(R) \, \psi_{\rm zero}^{\rm free}(R)\right)}\right]^d\quad ,
\label{removed}
\eea
Here, $
\psi_{\rm zero}^{\rm free}(r)=2^{d/2} \Gamma (\frac{d}{2}+1) \, I_{d/2}(r)/r^{d/2-1}$, and the classical solution behaves as
$
\Phi_{cl}(R)\sim \Phi_{\infty} \frac{K_{d/2-1}(R)}{R^{d/2-1}}$, as $R\to \infty$,
for some constant $\Phi_{\infty}$. Thus the $R\to\infty$ limit exists and is simple. Furthermore, from the virial theorem, 
$
\int_0^\infty dr\, r^{d-1}\, \psi_{\rm zero}^2(r)=\frac{\Gamma\left(\frac{d}{2}+1\right) S[\Phi_{cl}]}{\pi^{d/2}\, \left(\Phi_{cl}^{\prime\prime}(0)\right)^2} 
$,
so we arrive at the simple expression for the net contribution of this $l=1$ mode:
\bea
\left(\frac{S[\Phi_{cl}]}{2\pi}\right)^{d/2} \left(\frac{\det^\prime  \left[{\mathcal M}_{(l=1)}+m^2\right]}{\det \left[{\mathcal M}^{\rm free}_{(l=1)}+m^2\right]} \right)^{-1/2}=\left[\left(2\pi\right)^{d/2-1}\, \Phi_{\infty}\, \left | \Phi_{cl}^{\prime\prime}(0)\right |\right]^{d/2} \quad.
\label{zero-factor}
\eea
This expresses the zero mode factor solely in terms of the asymptotic properties of the classical solution $\Phi_{cl}$, which are already known numerically as part of the determination of $\Phi_{cl}$ and the fluctuation potential $V(r)$. No further computation is needed.

\section{Application 1: fluctuation determinant for false vacuum decay}

The phenomenon of nucleation drives first order phase transitions in many applications in physics. The semiclassical analysis of the rate of such a nucleation process was pioneered by Langer \cite{langer}, who identified a semiclassical saddle point solution that gives the dominant exponential contribution to the rate, with a prefactor to the exponential given by the quantum fluctuations about this classical solution. The nucleation rate is given by the quantum mechanical rate of decay of a metastable "false" vacuum, $\Phi_-$, into the "true" vacuum, $\Phi_+$. Decay proceeds by the nucleation of expanding bubbles of true vacuum within the metastable false vacuum \cite{langer,kobzarev,stone,coleman-fv,wipf,gorokhov}. 
The semiclassical prefactor requires computing the determinant of the differential operator associated with quantum fluctuations about the classical solution. This is a technically difficult problem, but it is ideally suited to the above results for determinants of radially separable operators, because the fluctuation operator is radial. The expression \eqn{dm-answer2} below can be applied to {\it any} metastable field potential $U(\Phi)$ \cite{dunnemin}, without relying on the "thin-wall" limit (the limit of degenerate vacua) . For definiteness, consider the specific (rescaled) quartic potential: $
U(\Phi)=\frac{1}{2}\Phi^2-\frac{1}{2}\Phi^3+\frac{\alpha}{8}\Phi^4
$, 
where $\alpha$ characterizes the shape of the potential [the thin-wall limit is $\alpha\to 1$]. The decay rate per unit volume and unit time is 
\bea
\gamma= \left(\frac{S_{\rm cl}[\phic]}{2\pi}\right)^2 \left|\frac{{\rm det}^\prime \left(-\Box + U^{\prime\prime}(\phic)\right)}{{\rm det}\left(-\Box +U^{\prime\prime}(\Phi_-)\right)}\right|^{-1/2}\, e^{-S_{\rm cl}[\phic]-\delta_{ct}S[\phic]}\quad ,
\label{rate}
\eea
where the prime on the determinant means that the zero modes (corresponding to translational invariance) are removed. Here $\phic$ is a classical solution known as the ``bounce'' solution \cite{coleman-fv}, defined below, and the prefactor terms in (\ref{rate}) correspond to quantum fluctuations about this bounce solution.  The second term in the exponent, $\delta_{ct}S[\phic]$, denotes the counterterms needed for renormalizing the classical action $S_{\rm cl}$. 
%\begin{figure}[h]
%\centerline{\includegraphics[scale=0.55]{fig2.eps}\hfill \includegraphics[scale=0.55]{fig4.eps}}
%\caption{Plots of the field potential [left], $U(\Phi)=\frac{1}{2}\Phi^2-\frac{1}{2}\Phi^3+\frac{\alpha}{8}\Phi^4$, for  $\alpha=0.6$, $0.7$, $0.8$ , $0.9$, $0.99.$ The vacua become degenerate in the "thin-wall" limit, $\alpha\to 1$. On the right are plots of the fluctuation potential $U^{\prime\prime}(\phic(r))$ for $\alpha= 0.5$,  $0.9$, $0.95$, $0.96$, $0.97$, $0.98$, $0.99$, with the binding well of the potential appearing farther to the right for increasing $\alpha$. Observe that as $\alpha\to 1$, the potential $U^{\prime\prime}(\phic(r))$ is localized at $r\sim \frac{1}{1-\alpha}$.}
%\label{fig2}
%\end{figure}
The first step in computing the false vacuum decay rate $\gamma$ is to find the classical bounce solution, $\phic(r)$, which is a radially symmetric stationary point of the classical Euclidean action,  interpolating between the false and true vacuum. 
The bounce $\phic(r)$ solves the nonlinear ordinary differential equation
\bea
-\phic^{\prime\prime} -\frac{3}{r}\phic^\prime +\phic-\frac{3}{2}\phic^2+\frac{\alpha}{2} \phic^3=0
\label{bounceeq}
\eea
with boundary conditions $
\phic^\prime(0)=0$, $
\phic(r)\to \Phi_-\equiv 0$, as $r\to\infty$. $\phic(r)$  must be computed numerically, for example by the method of shooting, adjusting the initial value $\phic(0)$ until the boundary conditions at $r=\infty$ are satisfied.
Given the bounce solution, $\phic(r)$, the corresponding radial fluctuation potential is 
\bea
U^{\prime\prime}(\phic(r))
= 1-3\phic(r)+\frac{3\alpha}{2}\phic^2(r)\quad .
\label{flucpot}
\eea

There are three different types of eigenvalue of the fluctuation operator, each having a different role physically and mathematically. (i) Negative Mode: The $l=0$ sector has a negative eigenvalue mode of the fluctuation operator, and is responsible for the instability leading to decay. 
This mode contributes a factor to the decay rate $\gamma$ related to the {\it absolute value} of the determinant of the $l=0$ fluctuation operator \cite{langer,coleman-fv}. (ii) Zero  Modes: In the $l=1$ sector, there is a four-fold degenerate zero eigenvalue of the fluctuation operator. 
Integrating over the corresponding collective coordinates produces the factors of $\frac{S_{\rm cl}}{2\pi}$ in (\ref{rate}). In computing the rate $\gamma$, we need the determinant of the fluctuation operator with the zero mode removed \cite{langer,coleman-fv,wipf}. This can be evaluated simply using the formula \eqn{zero-factor}, in terms of the asymptotic behavior of the classical bounce solution $\phic(r)$. (iii) Positive  Modes: For $l\geq 2$, the fluctuation operator has positive eigenvalues, each of degeneracy $(l+1)^2$. 
For each $l$, the associated radial determinant is computed numerically using \eqn{gyr}. 
\begin{figure}[ht]
\centerline{\includegraphics[scale=0.6]{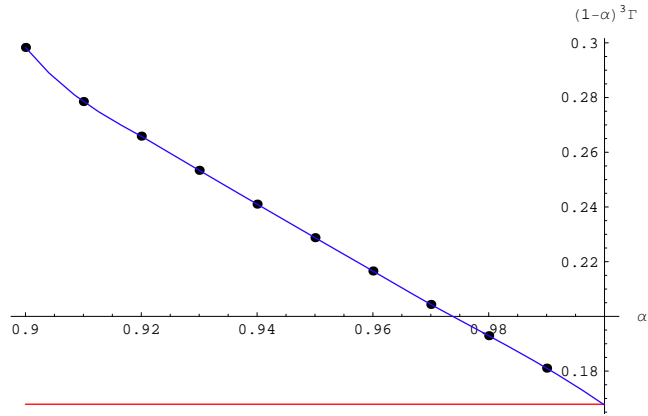}}
\caption{Comparison  as $\alpha\to 1$ of the exact numerical result [dots and solid blue line] from \eqn{dm-answer2} with the analytic thin-wall limit result $\frac{9}{32}\, \left[1-\frac{2\pi}{9\sqrt{3}}\right]\approx 0.16788..$ of \cite{kr}.}
\label{fig-alpha}
\end{figure}
But the sum over $l$ is divergent and the answer must be renormalized. Using the $\overline{{\rm MS}}$ scheme, the 
renormalized  effective action is \cite{dunnemin}:
\bea
\hskip-2cm \Gamma_{\overline{{\rm MS}}} &=& \frac{1}{2} \ln \left |T_{(0)}(\infty)\right |-2 \ln \left[\frac{\pi}{2} \Phi_\infty \left(\Phi_0-\frac{3}{2}\Phi_0^2+\frac{\alpha}{2}\Phi_0^3\right)\right]  \nn\\
\hskip -2cm && +\frac{1}{2} \sum_{l=2}^\infty (l+1)^2 \left\{ \ln \left(T_{(l)}(\infty)\right)-\frac{\frac{1}{2}\int_0^\infty dr \, r\, V(r)}{(l+1)}+\frac{\frac{1}{8}\int_0^\infty dr\, r^3\, V(V+2)}{(l+1)^3}\right\} \nn\\
\hskip -2cm&&  -\frac{3}{4}\int_0^\infty dr \, r\, V(r)  +\frac{1}{16} \int_0^\infty dr\, r^3\, V(V+2)\,\left(\frac{1}{2}-\gamma_E-\ln \frac{r}{2}\right)
\label{dm-answer2}
\eea
The first term is from the negative mode, the second from the zero mode(s) [using \eqn{zero-factor}], the third term from the numerical computation of the higher modes, and the last term is the renormalization counterterm contribution. The expression \eqn{dm-answer2} can be applied to {\it any} field potential $U(\Phi)$, without relying on the thin-wall limit. A comparison with the analytic thin-wall computation of Konoplich and Rubin \cite{kr} is plotted in Fig. \ref{fig-alpha}, and shows excellent agreement. This approach can also be extended to false vacuum decay in curved space, where the structure of the bounce solutions \cite{hackworth}, and the corresponding fluctuation equations \cite{dw}, is much richer.

\section{Application 2: determinant for \underline{massive} quarks in instanton background}
\label{idet-sec}

The fermion determinant in an instanton background is known in the massless (chiral) limit \cite{thooft}, and in the heavy quark limit \cite{nsvz}. Here we compute it for {\it any} $m$ \cite{idet}.
The relevant part of the renormalized effective action, 
$\tilde{\Gamma}^{S}_{\rm ren}(m)$, behaves as
\begin{eqnarray}
\hskip-2cm \tilde{\Gamma}^{S}_{\rm ren}(m)&=&
\begin{cases}
{\alpha(1/2)+\frac{1}{2}\left(\ln m+\gamma-\ln
2\right)m^2 +\dots \quad , \quad m\rightarrow 0 \cr 
\displaystyle {
 -\frac{\ln m}{6}-\frac{1}{75 m^2}-\frac{17}{735 m^4}+\frac{232}{2835 m^6}-\frac{7916}{148225 m^8}+\cdots \quad , \quad m\rightarrow \infty}}
\end{cases}
\label{masslimit}
\end{eqnarray}
where $\alpha(1/2)=  \simeq 0.145873$.  The small mass expansion \cite{thooft} is based on the known
massless propagators in an instanton background, while the large mass expansion  \cite{nsvz} follows most easily
from the Schwinger-DeWitt or heat kernel expansion.

Since the instanton is self-dual, we can simplify things slightly and work with the scalar Klein-Gordon operator. For a single $su(2)$ instanton, the spectral problem separates into partial waves \cite{thooft}, with radial Schr\"odinger-like operators (take isospin $\frac{1}{2}$)
\begin{equation}
\hskip -2cm {\cal H}_{(l,j)} \equiv \left[ - \frac{\partial^2}{\partial
r^2}-\frac{3}{r}\frac{\partial}{\partial
r}+\frac{4l(l+1)}{r^2}+\frac{4(j-l)(j+l+1)}{r^2+1}-\frac{3}{(r^2+1)^2}
\right]\, ,
\label{insth}
\end{equation}
Here $l=0, \frac{1}{2}, 1, \frac{3}{2}, \cdots\;$, and $j=| l
\pm \frac{1}{2}|$, and there is a degeneracy factor of
$(2l+1)(2j+1)$.
Since the "potential" involves $l$ and $j$, we cannot directly use the result \eqn{4d-result}.
In order to compare with 't Hooft's massless quark analysis \cite{thooft}, we use a Pauli-Villars regulator $\Lambda$, and define the renomalized effective action by  charge renormalization.
To extract the renormalized effective action we need to consider the $\Lambda\to\infty$ limit in conjunction with the infinite sum over $l$.  Split the partial wave sum into two parts:
\begin{eqnarray}
\Gamma_\Lambda^S(m)&=& \sum_{l=0,\frac{1}{2},\dots}^L \Gamma_{\Lambda, (l)}^S(m) + \sum_{l=L+\frac{1}{2}}^\infty \Gamma_{\Lambda, (l)}^S(m)
\label{actionsplit}
\end{eqnarray}
where $L$ is a large but finite integer. The first sum involves low partial wave modes, and the second sum involves the high partial wave modes. For the low partial wave modes, we can remove the regulator and evaluate the (finite) sum using the (numerical) Gel'fand-Yaglom result \eqn{gyr}. But this sum diverges quadratically with $L$. In the second sum in (\ref{actionsplit}) we cannot take the large
$L$ and large $\Lambda$ limits blindly, as each leads to a
divergence. Radial WKB  is a good approximation for the high $l$ modes. With WKB we can compute {\it analytically}  the large $\Lambda$ and
large $L$ divergences of the second sum in (\ref{actionsplit}), using the WKB approximation for the
corresponding determinants \cite{idet}:
\begin{eqnarray}
\hskip -2cm \sum_{l=L+\frac{1}{2}}^\infty \Gamma_{\Lambda, (l)}^S(A; m)&\sim& \frac{1}{6}\ln \Lambda+2 L^2 + 4 L-\left(\frac{1}{6}+\frac{m^2}{2}\right)\ln L \nonumber\\
&& +\left[\frac{127}{72}-\frac{1}{3}\ln 2+\frac{m^2}{2}-m^2 \ln 2+\frac{m^2}{2}\ln m \right]+O\left(\frac{1}{L}\right)
\label{div}
\end{eqnarray}
It is important to identify the physical role of the various terms in
(\ref{div}). The first term is the expected logarithmic charge renormalization counterterm.
The next three terms give
quadratic, linear and logarithmic divergences in $L$. These divergences cancel {\it exactly} against corresponding divergences in the first sum in (\ref{actionsplit}), which are
found numerically. 

\begin{figure}[tp]
\begin{centering}\psfrag{hor}{ $m$}
\psfrag{ver}{$\tilde\Gamma_{\rm ren}^S(m)$}
\includegraphics[scale=1.25]{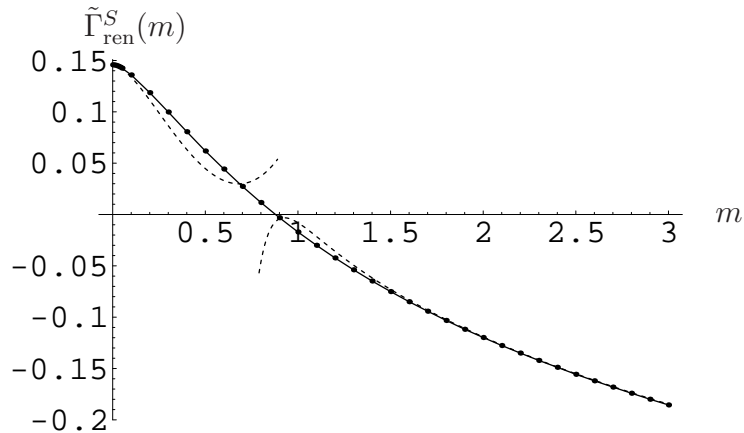}
\caption{Plot of our numerical results for $\tilde{\Gamma}^{S}_{\rm ren}(m)$
from (\protect{\ref{answer2}}), compared  with the analytic extreme small
and large mass limits
[dashed curves] from (\protect{\ref{masslimit}}).
The dots denote numerical data points from (\protect{\ref{answer2}}),
and the solid line is a fit through these points. \label{fig4}}
\end{centering}
\end{figure}

Combining the numerical results for  low partial wave modes with the radial WKB results for the high partial wave modes  the renormalized effective
action $\tilde{\Gamma}^S_{\rm ren}(m)$ is
\begin{eqnarray}
\hskip-1cm \tilde{\Gamma}^S_{\rm ren}(m)&=&\lim_{L\to \infty}\left\{\sum_{l=0,\frac{1}{2},\dots}^L (2l+1)(2l+2) P(l)
+2 L^2 + 4 L-\left(\frac{1}{6}+\frac{m^2}{2}\right)\ln L  \right.\nonumber \\
&&\left. +\left[\frac{127}{72}-\frac{1}{3}\ln 2+\frac{m^2}{2}-m^2
\ln 2+\frac{m^2}{2}\ln m \right]\right\}.
\label{answer2}
\end{eqnarray}
This is finite for any mass $m$.  Figure
\ref{fig4} shows these results for $\tilde{\Gamma}^{S}_{\rm ren}(m)$,
compared with the analytic
small and large mass expansions in (\ref{masslimit}). The agreement
is remarkable \cite{idet}. It is a highly nontrivial check on the
WKB computation that the correct mass dependence rises to interpolate between the large mass
and small mass regimes.
As an interesting {\it analytic} check, the formula (\ref{answer2}) also provides a 
simple computation \cite{idet} of 't Hooft's leading small mass result $\alpha(1/2)=-\frac{17}{72}-\frac{1}{6}\ln 2+\frac{1}{6}-2\zeta^\prime(-1)=0.145873... $.

\section{Concluding remarks}

In this talk I have reviewed some of the reasons for considering determinants of partial differential operators in quantum field theory. Mathematically, much is known for {\it ordinary} differential operators, especially through the work of Gel'fand-Yaglom \cite{gy,levit} and its extensions \cite{forman,kleinert,kirsten}, but considerably less is known for {\it partial} differential operators with non-trivial potentials. The new mathematical results \cite{dk} reported here are the formulas in \eqn{4d-result}, which provide simple expressions for the determinant of a radially separable partial differential operator of the form
$-\Delta+m^2+V(r)$. This generalizes the (Dirichlet) Gel'fand-Yaglom result
\eqn{gy1} to higher dimensions. 
These results lead to many direct applications in quantum field theory, where they extend the class of solvable fluctuation determinant problems away from the restrictive class of constant background fields, or one dimensional background fields, to the more general class of  separable higher dimensional background fields. This includes, for example, applications in quantum field theory to the study of quantum fluctuations in the presence of vortices, monopoles, sphalerons, instantons, domain walls (branes), etc ... 

A number of further generalizations could be made. First, it is practically  important  to include {\it
directly} the matrix structure that arises from Dirac-like differential operators and from non-abelian gauge degrees of freedom. This has been done in various ways for certain applications \cite{baacke2,carson,baacke3,strumia,burnier}, but a simpler unified formalism along the lines of the explicit formulas \eqn{4d-result} is probably possible, as in  one dimension \cite{waxman}. Second, a practical extension to finite temperature and density would be useful in various physical applications. Third, the biggest technical  challenge is to ask if the restriction of separability can be relaxed. A promising current approach is the numerical worldline loop method \cite{gies}.
Without symmetry, computing functional determinants analytically remains a difficult problem, and any progress will be interesting. 

\section*{Acknowledgments} I  thank the organizers, especially Mariano Del Olmo, for a stimulating conference, in a beautiful city. I acknowledge support from the DFG through the Mercator Guest Professor Program, and from the DOE through grant DE-FG02-92ER40716.


\vskip .2 cm

\begin{thebibliography}{12345}

 \bibitem{he}
  W.~Heisenberg and H.~Euler,
  ``Consequences of Dirac's theory of positrons,''
  Z.\ Phys.\  {\bf 98}, 714 (1936); English translation at
  [arXiv:physics/0605038].
  %%CITATION = ZEPYA,98,714;%
  
  \bibitem{schwinger}
  J.~S.~Schwinger,
  ``On gauge invariance and vacuum polarization,''
  Phys.\ Rev.\  {\bf 82}, 664 (1951).
  %%CITATION = PHRVA,82,664;%%
  
  \bibitem{gvd}
  G.~V.~Dunne,
  ``Heisenberg-Euler effective Lagrangians: Basics and extensions,'' in {\it Ian Kogan Memorial Collection, 'From Fields to Strings: Circumnavigating Theoretical Physics'},
M.~Shifman et al (ed.) vol. 1 pp 445-522 (World Scientific, 2005). 
  [arXiv:hep-th/0406216].
  %%CITATION = HEP-TH/0406216;%%
  
  \bibitem{ruijsenaars}
  S.~N.~M.~Ruijsenaars, ``On Barnes' Multiple Zeta and Gamma Functions'', Adv. Math. {\bf 156}, 107 (2000).
  
  \bibitem{camporesi} 
  R.~Camporesi,
  ``Harmonic analysis and propagators on homogeneous spaces,''
  Phys.\ Rept.\  {\bf 196}, 1 (1990);
  %%CITATION = PRPLC,196,1;%%
   I.~G.~Avramidi,
  ``Heat kernel and quantum gravity,''
  Lect.\ Notes Phys.\  {\bf M64}, 1 (2000).
  %%CITATION = LNPHA,M64,1;%%
  
  \bibitem{dowker}
J.~S.~Dowker,
  ``Functional determinants on spheres and sectors,''
  J.\ Math.\ Phys.\  {\bf 35}, 4989 (1994)
  [Erratum-ibid.\  {\bf 36}, 988 (1995)]
  [arXiv:hep-th/9312080];
  %%CITATION = JMAPA,35,4989;%%
J.~S.~Dowker and K.~Kirsten,
 ``The Barnes zeta-function, sphere determinants and Glaisher-
  Kinkelin-Bendersky constants,''
  Anal.\ Appl.\  {\bf 3}, 45 (2005)
  [arXiv:hep-th/0301143].
  %%CITATION = 00488,3,45;%%
  
  \bibitem{dhoker}
 E.~D'Hoker and D.~H.~Phong,
  ``On Determinants Of Laplacians On Riemann Surfaces,''
  Commun.\ Math.\ Phys.\  {\bf 104}, 537 (1986).
  %%CITATION = CMPHA,104,537;%%

\bibitem{sarnak}
  P.~Sarnak,
  ``Determinants of Laplacians'',
  Commun.\ Math.\ Phys.\ {\bf 110}, 113 (1987).
  
   \bibitem{thooft}
  G.~'t Hooft,
  ``Computation of the quantum effects due to a four-dimensional
  pseudoparticle,''
  Phys.\ Rev.\  D {\bf 14}, 3432 (1976)
  [Erratum-ibid.\  D {\bf 18}, 2199 (1978)];
  %%CITATION = PHRVA,D14,3432;%%
   R.~D.~Carlitz and D.~B.~Creamer,
  ``Light Quarks And Instantons,''
  Annals Phys.\  {\bf 118}, 429 (1979).
  %%CITATION = APNYA,118,429;%%
  
   \bibitem{osborn}
  H.~Osborn,
  ``Semiclassical Functional Integrals For Selfdual Gauge Fields,''
  Annals Phys.\  {\bf 135}, 373 (1981).
  %%CITATION = APNYA,135,373;%%
  
  \bibitem{gross}
  D.~J.~Gross, R.~D.~Pisarski and L.~G.~Yaffe,
  ``QCD And Instantons At Finite Temperature,''
  Rev.\ Mod.\ Phys.\  {\bf 53}, 43 (1981).
  %%CITATION = RMPHA,53,43;%%
  
  \bibitem{diakonov}
  D.~Diakonov, N.~Gromov, V.~Petrov and S.~Slizovskiy,
  ``Quantum weights of dyons and of instantons with non-trivial holonomy,''
  Phys.\ Rev.\  D {\bf 70}, 036003 (2004)
  [arXiv:hep-th/0404042].
  %%CITATION = PHRVA,D70,036003;%%
  
   \bibitem{seiler}
  H.~Hogreve, R.~Schrader and R.~Seiler,
  ``A Conjecture On The Spinor Functional Determinant,''
  Nucl.\ Phys.\  B {\bf 142}, 525 (1978);
  %%CITATION = NUPHA,B142,525;%%
   R.~Schrader and R.~Seiler,
  ``A Uniform Lower Bound On The Renormalized Scalar Euclidean Functional
  Determinant,''
  Commun.\ Math.\ Phys.\  {\bf 61}, 169 (1978).
  %%CITATION = CMPHA,61,169;%%
  D.~Brydges, J.~Frohlich and E.~Seiler,
  ``On The Construction Of Quantized Gauge Fields. I. General Results,''
  Annals Phys.\  {\bf 121}, 227 (1979).
  %%CITATION = APNYA,121,227;%%
   
  \bibitem{vafawitten}
  C.~Vafa and E.~Witten,
  ``Eigenvalue Inequalities For Fermions In Gauge Theories,''
  Commun.\ Math.\ Phys.\  {\bf 95}, 257 (1984).
  %%CITATION = CMPHA,95,257;%%
    
  \bibitem{fry}
  M.~P.~Fry,
  ``QED in inhomogeneous magnetic fields,''
  Phys.\ Rev.\  D {\bf 54}, 6444 (1996)
  [arXiv:hep-th/9606037];
  %%CITATION = PHRVA,D54,6444;%%
  ``Fermion determinant for general background gauge fields,''
  Phys.\ Rev.\  D {\bf 67}, 065017 (2003)
  [arXiv:hep-th/0301097].
  %%CITATION = PHRVA,D67,065017;%%
  
  \bibitem{vassilevich}
  D.~V.~Vassilevich,
  ``Heat kernel expansion: User's manual,''
  Phys.\ Rept.\  {\bf 388}, 279 (2003)
  [arXiv:hep-th/0306138].
  %%CITATION = PRPLC,388,279;%%
  
%  \bibitem{elizalde}
%  E.~Elizalde, S.~D.~Odintsov, A.~Romeo, A.~A.~Bytsenko and S.~Zerbini,
%  {\it Zeta regularization techniques with applications}, (World Scientific, Singapore, 1994).
  
  \bibitem{voros}
  A.~Voros,
`` Spectral Functions, Special Functions And Selberg Zeta Function'',
  Commun.\ Math.\ Phys.\  {\bf 110}, 439 (1987).
  %%CITATION = CMPHA,110,439;%%
  
  \bibitem{dhn}
  R.~F.~Dashen, B.~Hasslacher and A.~Neveu,
  ``Particle Spectrum In Model Field Theories From Semiclassical Functional
  Integral Techniques,''
  Phys.\ Rev.\  D {\bf 11}, 3424 (1975);
  %%CITATION = PHRVA,D11,3424;%%
  ``Semiclassical Bound States In An Asymptotically Free Theory,''
  Phys.\ Rev.\  D {\bf 12}, 2443 (1975).
  %%CITATION = PHRVA,D12,2443;%%
  
  \bibitem{thies}
  M.~Thies,
  ``From relativistic quantum fields to condensed matter and back again:
  Updating the Gross-Neveu phase diagram,''
  J.\ Phys.\ A  {\bf 39}, 12707 (2006)
  [arXiv:hep-th/0601049].
  %%CITATION = JPAGB,A39,12707;%%
  
  \bibitem{avan}
  L.~F.~Abbott and H.~J.~Schnitzer,
  ``Semiclassical Bound State Methods In Four-Dimensional Field Theory: Trace
  Identities, Mode Sums, And Renormalization For Scalar Theories,''
  Phys.\ Rev.\  D {\bf 14}, 1977 (1976);
  %%CITATION = PHRVA,D14,1977;%%
  J.~Avan and H.~J.~de Vega,
  ``Classical Solutions By Inverse Scattering Transformation In Any Number Of
  Dimensions. 1. The Gap Equation And The Effective Action,''
  Phys.\ Rev.\  D {\bf 29}, 2891 (1984).
  %%CITATION = PHRVA,D29,2891;%%
  
  \bibitem{gy}
I.~M.~Gelfand and A.~M.~Yaglom,
  ``Integration In Functional Spaces And It Applications In Quantum Physics,''
  J.\ Math.\ Phys.\  {\bf 1}, 48 (1960).
  %%CITATION = JMAPA,1,48;%%

\bibitem{levit}
S. Levit and U. Smilansky, ``A theorem on infinite products of eigenvalues of Sturm-Liouville type operators'', Proc. Am. Math. Soc. {\bf 65}, 299 (1977).

\bibitem{coleman}
S.~R.~Coleman,
``The Uses Of Instantons,'' in {\it Aspects of Symmetry}, (Cambridge University Press, 1985).
%{\it Lectures delivered at 1977 International School of Subnuclear Physics, Erice:
%The Whys of Subnuclear Physics}, Edited by A.  Zichichi, (Plenum Press, 1979).

\bibitem{forman}
R. Forman, `` Functional determinants and geometry '',
Invent. Math. {\bf 88}, 447 (1987); Erratum, {\it ibid} {\bf 108}, 453 (1992).


\bibitem{kappeler}
D.~Burghelea, L.~Friedlander and T.~Kappeler, ``On the determinant of elliptic differential and finite difference operators in vector bundles over $S^1$'', Commun. Math. Phys. {\bf 138}, 1 (1991). 

\bibitem{kleinert}
  H.~Kleinert,
  ``Path Integrals in Quantum Mechanics,  Statistics, Polymer Physics,
  and Financial Markets,'' (World Scientific, Singapore, 2004).

\bibitem{simon}
B.~Simon, ``Notes on infinite determinants of Hilbert space operators'', Adv. Math. {\bf 24}, 244 (1977).

\bibitem{kirsten}
  K.~Kirsten and A.~J.~McKane,
  ``Functional determinants by contour integration methods,''
  Annals Phys.\  {\bf 308}, 502 (2003)
  [arXiv:math-ph/0305010];
  %%CITATION = MATH-PH 0305010;%%
 ``Functional determinants for general Sturm-Liouville problems,''
 J.\ Phys.\ A {\bf 37}, 4649 (2004)
  [arXiv:math-ph/0403050].
  %%CITATION = MATH-PH 0403050;%%

\bibitem{dym} 
T.~Dreyfus and H.~Dym, ``Product formulas for the eigenvalues of a class of boundary value problems'', Duke Math. J. {\bf 45}, 15 (1978).

%\bibitem{abramowitz}
%M.~Abramowitz and I.~Stegun, {\it Handbook of Mathematical
%Functions}, (Dover, New York, 1965).

\bibitem{dk}
G.~V.~Dunne and K.~Kirsten,
  ``Functional determinants for radial operators,''
  J.\ Phys.\ A  {\bf 39}, 11915 (2006)
  [arXiv:hep-th/0607066].
  %%CITATION = JPAGB,A39,11915;%%

%\bibitem{schwinger}
%J.~Schwinger,
%``The Theory of Quantized Fields. VI'', Phys.\ Rev.\ {\bf 94}, 1362 (1954).

\bibitem{taylor}
J.~R.~Taylor,
{\it Scattering Theory}, (Wiley, New York, 1972).

\bibitem{kirstenbook}
K. Kirsten, {\it Spectral Functions in Mathematics and Physics}, (Chapman-Hall,  2001); 
 K.~Kirsten $\&$ P.~Loya,
  ``Computation of determinants using contour integrals,''
  arXiv:0707.3755 [hep-th].
  %%CITATION = ARXIV:0707.3755;%%

\bibitem{baacke}
  J.~Baacke and G.~Lavrelashvili,
  ``One-loop corrections to the metastable vacuum decay,''
  Phys.\ Rev.\ D {\bf 69}, 025009 (2004)
  [arXiv:hep-th/0307202].
  %%CITATION = HEP-TH 0307202;%%
  
  \bibitem{mckane}
  A.~J.~McKane and M.~B.~Tarlie,
  ``Regularisation of functional determinants using boundary perturbations,''
  J.\ Phys.\ A  {\bf 28}, 6931 (1995)
  [arXiv:cond-mat/9509126].
  %%CITATION = JPAGB,A28,6931;%%
  
   \bibitem{dunnemin}
 G.~V.~Dunne and H.~Min,
  ``Beyond the thin-wall approximation: Precise numerical computation of prefactors in false vacuum decay,''
  Phys.\ Rev.\ D {\bf 72}, 125004 (2005)
  [arXiv:hep-th/0511156].
  %%CITATION = HEP-TH 0511156;%%

  \bibitem{langer}
  J.~S.~Langer,
  ``Theory Of The Condensation Point,''
  Annals Phys.\  {\bf 41}, 108 (1967).
  %%CITATION = APNYA,41,108;%%

\bibitem{kobzarev}
  M.~B.~Voloshin, I.~Y.~Kobzarev, and L.~B.~Okun,
  ``Bubbles In Metastable Vacuum,''
  Sov.\ J.\ Nucl.\ Phys.\  {\bf 20}, 644 (1975)
  [Yad.\ Fiz.\  {\bf 20}, 1229 (1974)].
  %%CITATION = SJNCA,20,644;%%

  \bibitem{stone}
  M.~Stone,
   ``The Lifetime And Decay Of 'Excited Vacuum' States Of A Field Theory
  Associated With Nonabsolute Minima Of Its Effective Potential,''
  Phys.\ Rev.\ D {\bf 14}, 3568 (1976);
  %%CITATION = PHRVA,D14,3568;%%
 ``Semiclassical Methods For Unstable States,''
  Phys.\ Lett.\ B {\bf 67}, 186 (1977).
  %%CITATION = PHLTA,B67,186;%%
  
  \bibitem{coleman-fv}
  S.~R.~Coleman,
 ``The Fate Of The False Vacuum. 1. Semiclassical Theory,''
  Phys.\ Rev.\ D {\bf 15}, 2929 (1977)
  [Erratum-ibid.\ D {\bf 16}, 1248 (1977)];
  %%CITATION = PHRVA,D15,2929;%%
  C.~G.~Callan and S.~R.~Coleman,
  ``The Fate Of The False Vacuum. 2. First Quantum Corrections,''
  Phys.\ Rev.\ D {\bf 16}, 1762 (1977).
  %%CITATION = PHRVA,D16,1762;%%

%  \bibitem{gervais}
% J.~L.~Gervais and B.~Sakita,
% ``WKB Wave Function For Systems With Many Degrees Of Freedom: A Unified  View
%  Of Solitons And Pseudoparticles,''
%  Phys.\ Rev.\ D {\bf 16}, 3507 (1977).
%  %%CITATION = PHRVA,D16,3507;%%
  
  \bibitem{wipf}
  A.~W.~Wipf,
  ``Tunnel Determinants,''
  Nucl.\ Phys.\  B {\bf 269}, 24 (1986).
  %%CITATION = NUPHA,B269,24;%%
  
  \bibitem{gorokhov}
  D.~A.~Gorokhov and G.~Blatter, ``Metastability of $(d+n)$-dimensional elastic manifolds'', Phys. Rev. B {\bf 58}, 5486 (1998).
  
  \bibitem{kr}
  R.~V.~Konoplich and S.~G.~Rubin,
  ``Quantum Corrections To Nontrivial Classical Solutions In Phi**4 Theory,''
  Yad.\ Fiz.\  {\bf 37}, 1330 (1983).
  %%CITATION = YAFIA,37,1330;%%
%  `` Decay Probability For Metastable Vacuum In Scalar Theory'',
%  Yad.\ Fiz.\  {\bf 42}, 1282 (1985).
%  %%CITATION = YAFIA,42,1282;%
  
  \bibitem{hackworth}
  J.~C.~Hackworth and E.~J.~Weinberg,
  ``Oscillating bounce solutions and vacuum tunneling in de Sitter
  spacetime,''
  Phys.\ Rev.\  D {\bf 71}, 044014 (2005)
  [arXiv:hep-th/0410142].
  %%CITATION = PHRVA,D71,044014;%%

  \bibitem{dw}
  G.~V.~Dunne and Q.~h.~Wang,
  ``Fluctuations about cosmological instantons,''
  Phys.\ Rev.\  D {\bf 74}, 024018 (2006)
  [arXiv:hep-th/0605176].
  %%CITATION = PHRVA,D74,024018;%%
  
  \bibitem{nsvz}
  V.~A.~Novikov, M.~A.~Shifman, A.~I.~Vainshtein and V.~I.~Zakharov,
  ``Calculations In External Fields In Quantum Chromodynamics. Technical
  Review,''
  Fortsch.\ Phys.\  {\bf 32}, 585 (1984).
  %%CITATION = FPYKA,32,585;%%

\bibitem{idet}
  G.~V.~Dunne, J.~Hur, C.~Lee and H.~Min,
   ``Instanton determinant with arbitrary quark mass: WKB phase-shift method and
  derivative expansion,''
  Phys.\ Lett.\ B {\bf 600}, 302 (2004)
  [arXiv:hep-th/0407222];
  %%CITATION = HEP-TH 0407222;%%
  ``Precise quark mass dependence of instanton determinant,''
  Phys.\ Rev.\ Lett.\  {\bf 94}, 072001 (2005)
  [arXiv:hep-th/0410190];
  %%CITATION = HEP-TH 0410190;%%
``Calculation of QCD instanton determinant with arbitrary mass,''
  Phys.\ Rev.\ D {\bf 71}, 085019 (2005)
  [arXiv:hep-th/0502087].
  %%CITATION = HEP-TH 0502087;%%


\bibitem{baacke2}
  J.~Baacke,
   ``Numerical Evaluation Of The One Loop Effective Action In Static Backgrounds
   With Spherical Symmetry,''
  %
  Z.\ Phys.\ C {\bf 47}, 263 (1990);
  %%CITATION = ZEPYA,C47,263;%%
 ``The effective action of a spin 1/2 field in the background of a chiral soliton'',   Z. Phys. C {\bf 53}, 407 (1992).

 \bibitem{carson}
  L.~Carson, X.~Li, L.~D.~McLerran and R.~T.~Wang,
   ``Exact Computation Of The Small Fluctuation Determinant Around A Sphaleron,''
  Phys.\ Rev.\ D {\bf 42}, 2127 (1990).
  %%CITATION = PHRVA,D42,2127;%%

  \bibitem{baacke3}
  J.~Baacke and S.~Junker,
   ``Quantum fluctuations around the electroweak sphaleron,''
  Phys.\ Rev.\ D {\bf 49}, 2055 (1994)
  [arXiv:hep-ph/9308310];
  %%CITATION = HEP-PH 9308310;%%
   ``Quantum fluctuations of the electroweak sphaleron: Erratum and addendum,''
  Phys.\ Rev.\ D {\bf 50}, 4227 (1994)
  [arXiv:hep-th/9402078].
  %%CITATION = HEP-TH 9402078;%%

\bibitem{strumia}
  G.~Isidori, G.~Ridolfi and A.~Strumia,
   ``On the metastability of the standard model vacuum,''
  Nucl.\ Phys.\ B {\bf 609}, 387 (2001)
  [arXiv:hep-ph/0104016].
  %%CITATION = HEP-PH 0104016;%%

  \bibitem{burnier}
  Y.~Burnier and M.~Shaposhnikov,
  ``One-loop fermionic corrections to the instanton transition in two dimensional chiral Higgs model,''
  Phys.\ Rev.\ D {\bf 72}, 065011 (2005)
  [arXiv:hep-ph/0507130].
  %%CITATION = HEP-PH 0507130;%%
  
  \bibitem{waxman}
  D.~Waxman,
  ``The Fredholm Determinant of A Dirac Operator,''
  Annals Phys.\  {\bf 231}, 256 (1994).
  %%CITATION = APNYA,231,256;%%
  
  \bibitem{gies}
  H.~Gies and K.~Langfeld,
  ``Quantum diffusion of magnetic fields in a numerical worldline approach,''
  Nucl.\ Phys.\  B {\bf 613}, 353 (2001)
  [arXiv:hep-ph/0102185];
  %%CITATION = NUPHA,B613,353;%%
``Loops and loop clouds: A numerical approach to the worldline formalism  in
  QED,''
  Int.\ J.\ Mod.\ Phys.\  A {\bf 17}, 966 (2002)
  [arXiv:hep-ph/0112198].
  %%CITATION = IMPAE,A17,966;%%

  
   

  \end{thebibliography}
\end{document}